\documentclass[11pt, a4paper]{article}
\usepackage[utf8]{inputenc}
\usepackage[left=2cm, right=2cm, bottom=2cm, top=2cm]{geometry}
\usepackage{xcolor}
\usepackage{bm}
\usepackage{amsmath, amsfonts}
\usepackage{siunitx}
\DeclareSIUnit{\angstrom}{\textup{\AA}}
\usepackage[hyphens]{url}
\usepackage{booktabs}
\usepackage{graphicx}
\usepackage{caption}
\usepackage{subcaption}
\usepackage{csquotes}
\usepackage{xparse}
\definecolor{amethyst}{rgb}{0.6, 0.4, 0.8}
\definecolor{revcol}{rgb}{0,0,1}
\usepackage[square, numbers, comma, sort&compress]{natbib}
\usepackage[colorlinks, citecolor=blue, linkcolor=teal, urlcolor=amethyst]{hyperref}

\NewDocumentCommand{\qnameref}{sm}{%
  \enquote{%
    \IfBooleanTF{#1}{\nameref*{#2}}{\nameref{#2}}%
  }%
}

\usepackage[capitalise,noabbrev,nameinlink]{cleveref}
\crefname{sfig}{Supplementary Figure}{Supplementary Figures}
\crefname{stab}{Supplementary Table}{Supplementary Tables}
\usepackage{authblk}

\title{Protein language models trained on multiple sequence alignments learn phylogenetic relationships}
\author{Umberto Lupo\textsuperscript{1,2,*}, Damiano Sgarbossa\textsuperscript{1,2}, Anne-Florence Bitbol\textsuperscript{1,2,*}}
\affil{\textbf{1} Institute of Bioengineering, School of Life Sciences, École Polytechnique Fédérale de Lausanne (EPFL), CH-1015 Lausanne, Switzerland\\
\textbf{2} SIB Swiss Institute of Bioinformatics, CH-1015 Lausanne, Switzerland\\
* Corresponding authors: \href{mailto:umberto.lupo@epfl.ch}{umberto.lupo@epfl.ch}, \href{mailto:anne-florence.bitbol@epfl.ch}{anne-florence.bitbol@epfl.ch}}
\date{}

\begin{document}

\maketitle

\begin{abstract}
Self-supervised neural language models with attention have recently been applied to biological sequence data, advancing structure, function and mutational effect prediction. Some protein language models, including MSA Transformer and AlphaFold's EvoFormer, take multiple sequence alignments (MSAs) of evolutionarily related proteins as inputs. Simple combinations of MSA Transformer's row attentions have led to state-of-the-art unsupervised structural contact prediction. We demonstrate that similarly simple, and universal, combinations of MSA Transformer's column attentions strongly correlate with Hamming distances between sequences in MSAs. Therefore, MSA-based language models encode detailed phylogenetic relationships. We further show that these models can separate coevolutionary signals encoding functional and structural constraints from phylogenetic correlations reflecting historical contingency. To assess this, we generate synthetic MSAs, either without or with phylogeny, from Potts models trained on natural MSAs. We find that unsupervised contact prediction is substantially more resilient to phylogenetic noise when using MSA Transformer versus inferred Potts models.
\end{abstract}

\section*{Introduction}

The explosion of available biological sequence data has led to multiple computational approaches aiming to infer three-dimensional structure, biological function, fitness, and evolutionary history of proteins from sequence data~\cite{deJuan13,Cocco18}.
Recently, self-supervised deep learning models based on natural language processing methods, especially attention~\cite{Bahdanau14} and transformers~\cite{Vaswani17}, have been trained on large ensembles of protein sequences by means of the masked language modeling objective of filling in masked amino acids in a sequence, given the surrounding ones \cite{ElnaggarPreprint,Rives21,rao2021transformer,Choromanski20,Madani20,Madani21}.
These models, which capture long-range dependencies, learn rich representations of protein sequences, and can be employed for multiple tasks.
In particular, they can predict structural contacts from single sequences in an unsupervised way~\cite{rao2021transformer}, presumably by transferring knowledge from their large training set~\cite{Bhattacharya22}.
Neural network architectures based on attention are also employed in the Evoformer blocks in AlphaFold~\cite{Jumper21}, as well as in RoseTTAFold~\cite{Baek21} and RGN2~\cite{Chowdhury21}, and they contributed to the recent breakthrough in the supervised prediction of protein structure.

Protein sequences can be classified in families of homologous proteins, that descend from an ancestral protein and share a similar structure and function.
Analyzing multiple sequence alignments (MSAs) of homologous proteins thus provides substantial information about functional and structural constraints~\cite{deJuan13}.
The statistics of MSA columns, representing amino-acid sites, allow to identify functional residues that are conserved during evolution, and correlations of amino-acid usage between columns contain key information about functional sectors and structural contacts~\cite{Casari95,Socolich05,Dunn08,Halabi09}.
Indeed, through the course of evolution, contacting amino acids need to maintain their physico-chemical complementarity, which leads to correlated amino-acid usages at these sites: this is known as coevolution.
Potts models, also known as Direct Coupling Analysis (DCA), are pairwise maximum entropy models trained to match the empirical one- and two-body frequencies of amino acids observed in the columns of an MSA of homologous proteins~\cite{Lapedes99,Weigt09,Marks11,Morcos11,Sulkowska12,Ekeberg13,Ekeberg14,Figliuzzi18,Cocco18}.
They capture the coevolution of contacting amino acids, and they provided state-of-the-art unsupervised predictions of structural contacts before the advent of protein language models.
Note that coevolutionary signal also aids supervised contact prediction~\cite{Abriata18}. 

While most protein language neural networks take individual amino-acid sequences as inputs, some others have been trained to perform inference from MSAs of evolutionarily related sequences.
This second class of networks includes MSA Transformer~\cite{rao2021msa} and the Evoformer blocks in AlphaFold~\cite{Jumper21}, both of which interleave row (i.e.\ per-sequence) attention with column (i.e.\ per-site) attention.
Such an architecture is conceptually extremely attractive because it can incorporate coevolution in the framework of deep learning models using attention.
In the case of MSA Transformer, simple combinations of the model's row attention heads have led to state-of-the-art unsupervised structural contact predictions, outperforming both language models trained on individual sequences and Potts models~\cite{rao2021msa}.
Beyond structure prediction, MSA Transformer is also able to predict mutational effects~\cite{meier2021language,hie2022evolutionary} and to capture fitness landscapes~\cite{HawkinsPreprint}.
In addition to coevolutionary signal caused by structural and functional constraints, MSAs feature correlations that directly stem from the common ancestry of homologous proteins, i.e.\ from phylogeny.
Does MSA Transformer learn to identify phylogenetic relationships between sequences, which are a key aspect of the MSA data structure?

Here, we show that simple, and universal, combinations of MSA Transformer's column attention heads, computed on a given MSA, strongly correlate with the Hamming distances between sequences in that MSA.
This demonstrates that MSA Transformer encodes detailed phylogenetic relationships.
Is MSA Transformer able to separate coevolutionary signals encoding functional and structural constraints from phylogenetic correlations arising from historical contingency?
To address this question, we generate controlled synthetic MSAs from Potts models trained on natural MSAs, either without or with phylogeny.
For this, we perform Metropolis Monte Carlo sampling under the Potts Hamiltonians, either at equilibrium or along phylogenetic trees inferred from the natural MSAs.
Using the top Potts model couplings as proxies for structural contacts, we demonstrate that unsupervised contact prediction via MSA Transformer is substantially more resilient to phylogenetic noise than contact prediction using inferred Potts models.

\section*{Results}

\subsection*{Column attention heads capture Hamming distances in separate MSAs}
\label{subsec:results_separate_hamming}

\begin{figure}[htbp]
    \centering
    \includegraphics{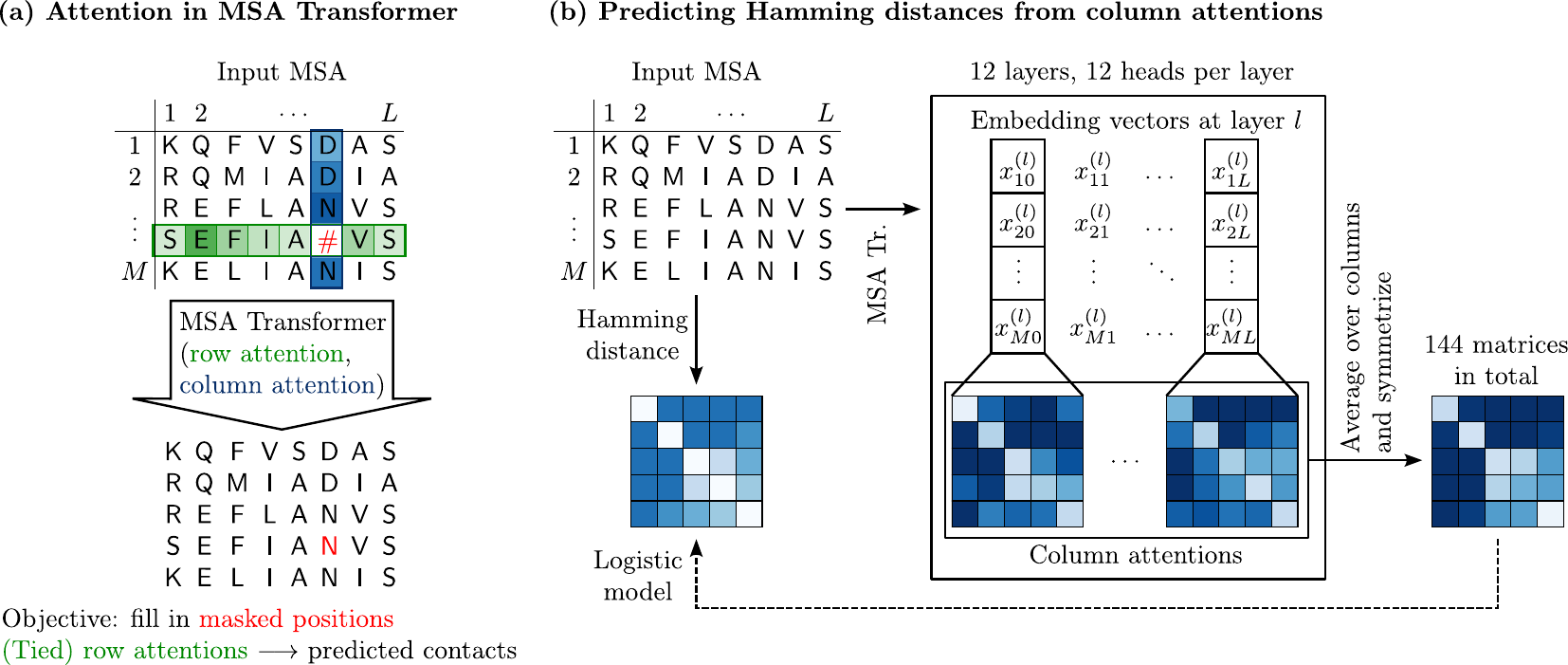}
    \caption{\textbf{MSA Transformer: column attentions and Hamming distances.} \textbf{(a)} MSA Transformer is trained using the masked language modeling objective of filling in randomly masked residue positions in MSAs.
    For each residue position in an input MSA, it assigns attention scores to all residue positions in the same row (sequence) and column (site) in the MSA.
    These computations are performed by 12 independent row/column attention heads in each of 12 successive layers of the network.
    \textbf{(b)} Our approach for Hamming distance matrix prediction from the column attentions computed by the trained MSA Transformer model, using a natural MSA as input.
    For each $i = 1, \ldots, M$, $j = 0, \ldots, L$ and $l = 1, \ldots, 12$, the embedding vector $x_{ij}^{(l)}$ is the $i$-th row of the matrix $X_{j}^{(l)}$ defined in the main text, and the column attentions are computed according to \cref{eq:query_key_values,eq:col_attn}.}
    \label{fig:msa_transformer_and_hamming_regr_schematics}
\end{figure}

\begin{figure}[htbp]
    \centering
    \begin{subfigure}[h]{0.73\textwidth}
        \centering
        \includegraphics[width=\textwidth]{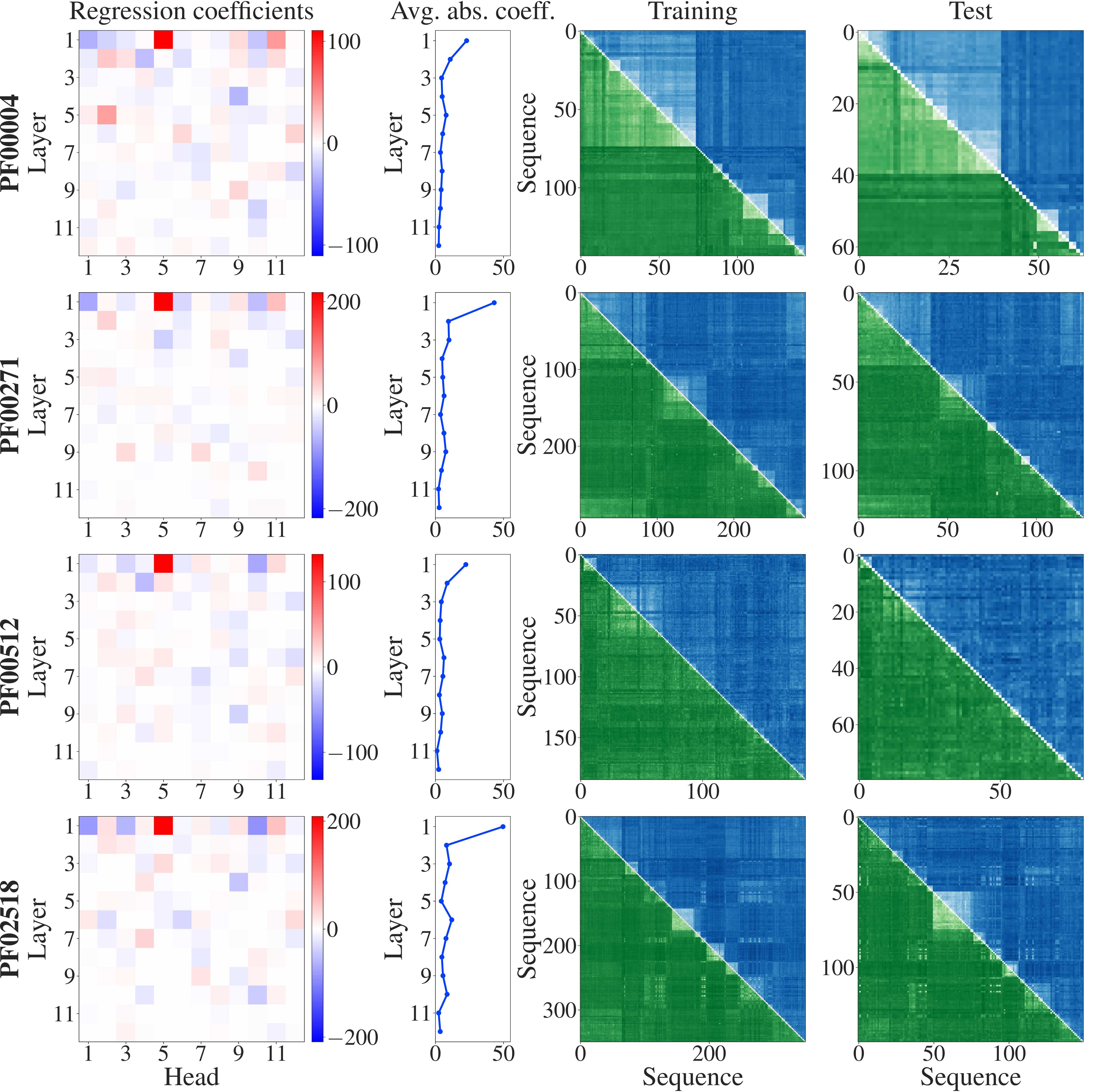}
        \caption{\label{subfig:mean_cols_regr_plots}}
    \end{subfigure}
    \hfill
    \begin{subfigure}[h]{0.25\textwidth}
        \centering
        \begin{tabular}{ll}
            \toprule
            \multicolumn{1}{c}{Family}  &  \multicolumn{1}{c}{$R^2$} \\
            \midrule
            PF00004 &       0.97 \\
            PF00005 &       0.99 \\
            PF00041 &       0.98 \\
            PF00072 &       0.99 \\
            PF00076 &       0.98 \\
            PF00096 &       0.94 \\
            PF00153 &       0.95 \\
            PF00271 &       0.94 \\
            PF00397 &       0.84 \\
            PF00512 &       0.94 \\
            PF00595 &       0.98 \\
            PF01535 &       0.86 \\
            PF02518 &       0.92 \\
            PF07679 &       0.99 \\
            PF13354 &       0.99 \\
            \bottomrule
        \end{tabular}
        \caption{\label{subfig:mean_cols_regr_table}}
    \end{subfigure}
    \caption{\textbf{Fitting logistic models to predict Hamming distances separately in each MSA.} The column-wise means of MSA Transformer's column attention heads are used to predict normalised Hamming distances as probabilities in a logistic model.
    Each MSA is randomly split into a training set comprising $70\%$ of its sequences and a test set composed of the remaining sequences.
    For each MSA, a logistic model is trained on all pairwise distances in the training set. \subref{subfig:mean_cols_regr_plots} Regression coefficients are shown for each layer and attention head (first column), as well as their absolute values averaged over heads for each layer (second column).
    For four example MSAs, ground truth Hamming distances are shown in the upper triangle (blue) and predicted Hamming distances in the lower triangle and diagonal (green), for the training and test sets (third and fourth columns).
    Darker shades correspond to larger Hamming distances. \subref{subfig:mean_cols_regr_table} $R^2$ coefficients of determination for the predictions by each fitted model on its respective test set.}
    \label{fig:mean_cols_regr_results}
\end{figure}

We first considered separately each of 15 different Pfam seed MSAs (see \qnameref{subsec:datasets} and \cref{tab:MSA}), corresponding to distinct protein families, and asked whether MSA Transformer has learned to encode phylogenetic relationships between sequences in its attention layers.
To test this, we split each MSA randomly into a training and a test set, and train a logistic model [\cref{eq:logistic_model_defn,eq:logistic_model_loss}] based on the column-wise means of MSA Transformer's column attention heads on all pairwise Hamming distances in the training set -- see \cref{fig:msa_transformer_and_hamming_regr_schematics} for a schematic, and \qnameref{subsec:methods_supervised_hamming} for details.
\cref{fig:mean_cols_regr_results} shows the results of fitting these specialized logistic models.

For all alignments considered, large regression coefficients concentrate in early layers in the network, and single out some specific heads consistently across different MSAs -- see \cref{subfig:mean_cols_regr_plots}, first and second columns, for results on four example MSAs.
These logistic models reproduce the Hamming distances in the training set very well, and successfully predict those in the test set -- see \cref{subfig:mean_cols_regr_plots}, third and fourth columns, for results on four example MSAs.
Note that the block structures visible in the Hamming distance matrices, and well reproduced by our models, come from the phylogenetic ordering of sequences in our seed MSAs, see \qnameref{subsec:datasets}.
Quantitatively, in all the MSAs studied, the coefficients of determination ($R^2$) computed on the test sets are above $0.84$ in all our MSAs -- see \cref{subfig:mean_cols_regr_table}.

\begin{figure}[htbp]
    \centering
    \includegraphics[width=0.3\textwidth]{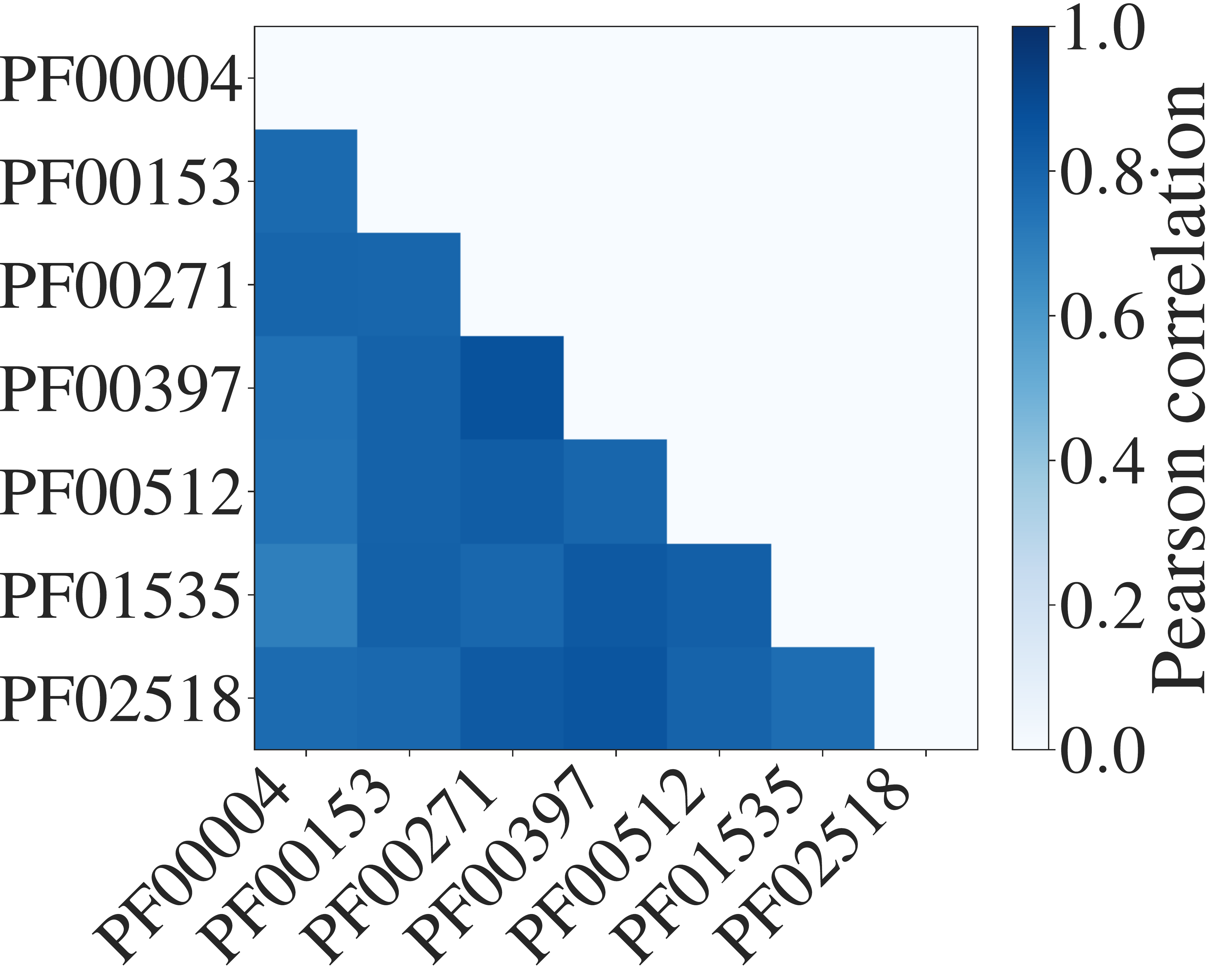}
    \caption{\textbf{Pearson correlations between regression coefficients in larger MSAs.}  Sufficiently deep ($\geq 100$ sequences) and long ($\geq 30$ residues) MSAs are considered (mean/min/max Pearson correlations: $0.80$/$0.69$/$0.87$).}
    \label{fig:mean_cols_regr_coeffs_pearson_corr}
\end{figure}

A striking result from our analysis is that the regression coefficients appear to be similar across MSAs -- see \cref{subfig:mean_cols_regr_plots}, first column.
To quantify this, we computed the Pearson correlations between the regression coefficients learnt on the larger seed MSAs.
\cref{fig:mean_cols_regr_coeffs_pearson_corr} demonstrates that regression coefficients are indeed highly correlated across these MSAs.

\clearpage

\subsection*{MSA Transformer learns a universal representation of Hamming distances}
\label{subsec:results_common_hamming}

\begin{figure}[htbp]
    \centering
    \includegraphics[width=\textwidth]{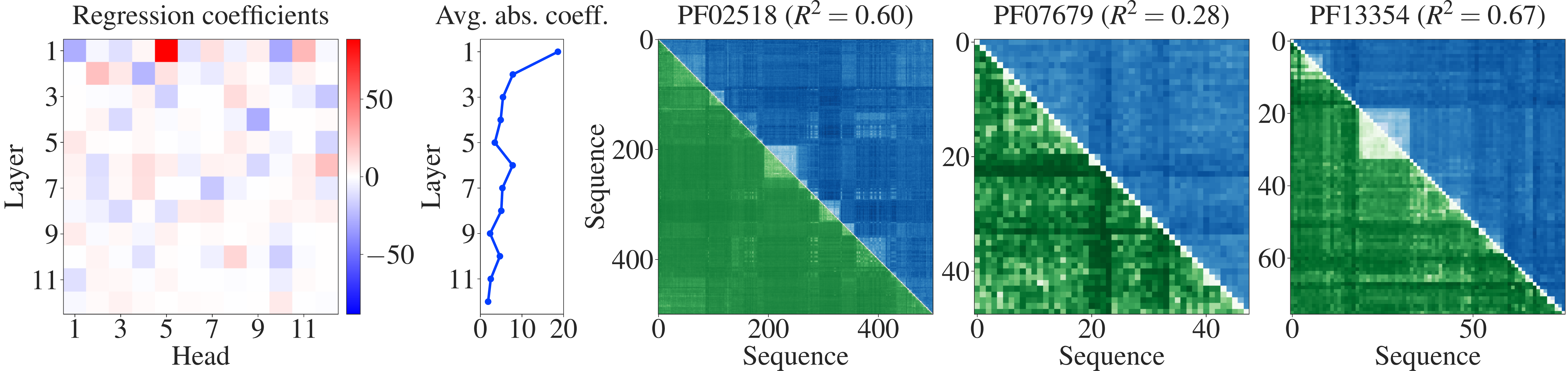}
    \caption{\textbf{Fitting a single logistic model to predict Hamming distances.}
    Our collection of $15$ MSAs is split into a training set comprising $12$ of them and a test set composed of the remaining $3$.
    A logistic regression is trained on all pairwise distances within each MSA in the training set.
    Regression coefficients (first panel) and their absolute values averaged over heads for each layer (second panel) are shown as in \cref{fig:mean_cols_regr_results}.
    For the three test MSAs, ground truth Hamming distances are shown in the upper triangle (blue) and predicted Hamming distances in the lower triangle and diagonal (green), also as in \cref{fig:mean_cols_regr_results} (last three panels).
    We further report the $R^2$ coefficients of determination for the regressions on these test MSAs -- see also \cref{tab:mean_cols_regr_results_many_msas_fit_quality}.}
    \label{fig:mean_cols_regr_results_many_msas}
\end{figure}

Given the substantial similarities between our models trained separately on different MSAs, we next asked whether a common model across MSAs could capture Hamming distances within generic MSAs.
To address this question, we trained a single logistic model, based on the column-wise means of MSA Transformer’s column attention heads, on all pairwise distances within each of the first 12 of our seed MSAs.
We assessed its ability to predict Hamming distances in the remaining 3 seed MSAs, which thus correspond to entirely different Pfam families from those in the training set.
\cref{fig:mean_cols_regr_results_many_msas} shows the coefficients of this regression (first and second panels), as well as comparisons between predictions and ground truth values for the Hamming distances within the three test MSAs (last three panels).
We observe that large regression coefficients again concentrate in the early layers of the model, but somewhat less than in individual models.
Furthermore, the common model captures well the main features of the Hamming distance matrices in test MSAs.

In \cref{tab:mean_cols_regr_results_many_msas_fit_quality}, we quantify the quality of fit for this model on all our MSAs.
In all cases, we find very high Pearson correlation between the predicted distances and the ground truth Hamming distances.
Furthermore, the median value of the $R^2$ coefficient of determination is 0.6, confirming the good quality of fit.
In the three shortest and the two shallowest MSAs, the model performs below this median, while all MSAs for which $R^2$ is above median satisfy $M \geq 52$ and $L \geq 67$.
We also compute, for each MSA, the slope of the linear fit when regressing the ground truth Hamming distances on the distances predicted by the model.
MSA depth is highly correlated with the value of this slope (Pearson $r \approx 0.95$).
This bias may be explained by the under-representation in the training set of Hamming distances and attention values from shallower MSAs, as their number is quadratic in MSA depth.

Ref.~\cite{rao2021msa} showed that some column attention matrices, summed along one of their dimensions, correlate with phylogenetic sequence weights (see \qnameref{subsec:methods_supervised_hamming}). This indicates that the model is, in part, attending to maximally diverse sequences.
Our study demonstrates that MSA Transformer actually learns pairwise phylogenetic relationships between sequences, beyond these aggregate phylogenetic sequence weights.
It also suggests an additional mechanism by which the model may be attending to these relationships, focusing on similarity instead of diversity.
Indeed, while our regression coefficients with positive sign in \cref{fig:mean_cols_regr_results_many_msas} are associated with (average) attentions that are positively correlated with the Hamming distances, we also find several coefficients with large negative values.
They indicate the existence of important negative correlations: in those heads, the model is actually attending to pairs of similar sequences.
Besides, comparing our \cref{subfig:mean_cols_regr_plots,fig:mean_cols_regr_results_many_msas} with Fig.~5 in Ref.~\cite{rao2021msa} shows that different attention heads are important in our study versus in the analysis of Ref.~\cite[Sec.\ 5.1]{rao2021msa}.
Specifically, here we find that the fifth attention head in the first layer in the network is associated with the largest positive regression coefficient, while the sixth one was most important there.
Moreover, still focusing on the first layer of the network, the other most prominent heads here were not significant there.
MSA Transformer's ability to focus on similarity may also explain why its performance at predicting mutational effects can decrease significantly when using MSAs which include a duplicate of the query sequence~\cite[Supplementary Fig.~9 and Table~10]{meier2021language}: in these cases, the model predicts masked tokens with very high confidence using information from the duplicate sequence.

How much does the ability of MSA Transformer to capture phylogenetic relationships arise from its training?
To address this question, we trained a common logistic model as above to predict Hamming distances, but using column attention values computed from a randomly re-initialized version of the MSA Transformer network.
We used the same protocol as in MSA Transformer's original pre-training to randomly initialize the entries of the network's row- and column-attention weight matrices $W_\mathrm{\mathrm{Q}}^{(l, h)}$, $W_\mathrm{\mathrm{K}}^{(l, h)}$ and $W_\mathrm{\mathrm{V}}^{(l, h)}$ (see \qnameref{subsec:MSA-Tr_attn}), as well as the entries of the matrix used to embed input tokens, the weights in the feed-forward layers, and the positional encodings.
Specifically, we sampled these entries (with the exception of bias terms and of the embedding vector for the padding token, which were set to zero) from a Gaussian distribution with mean $0$ and standard deviation $0.02$.
The results obtained in this case for our regression task are reported in \cref{tab:mean_cols_regr_results_many_msas_random_init_fit_quality}.
They demonstrate that, although random initialization can yield better performance than random guessing (which may partly be explained by Gordon's Theorem \cite{gordon1988milman}), the trained MSA Transformer gives vastly superior results.
This confirms that the masked language modeling pre-training has driven it towards precisely encoding distances between sequences.

For each layer and attention head in the network, MSA Transformer computes one matrix of column attention values per site -- see \cref{eq:A^lh}.
This is in contrast with row attention, which is tied (see \qnameref{subsec:MSA-Tr_attn}).
Our results are more surprising that they would be if the model's column attentions were also tied.
Indeed, during pre-training, by tuning its row-attention weight matrices to achieve optimal tied attention, MSA Transformer discovers covariance between MSA sites in early layers, and
covariance between MSA sequences is related to Hamming distance.

Finally, to explore the contribution of each column to performance in our regression task, we employed our common logistic model (trained on the means of column attention matrices) to predict Hamming distances using column attentions from individual sites.
We find that the most highly conserved sites (corresponding to columns with low entropy) lead to predictions whose errors have among the smallest standard deviations -- see \cref{tab:low_entropy_table}.
Note that we focused on standard deviations to mitigate the biases of the common logistic model (see above).
This indicates that highly conserved sites lead to more stable predictions.

\subsection*{MSA Transformer efficiently disentangles correlations from contacts and phylogeny}

\begin{figure}[htbp]
    \centering
    \includegraphics{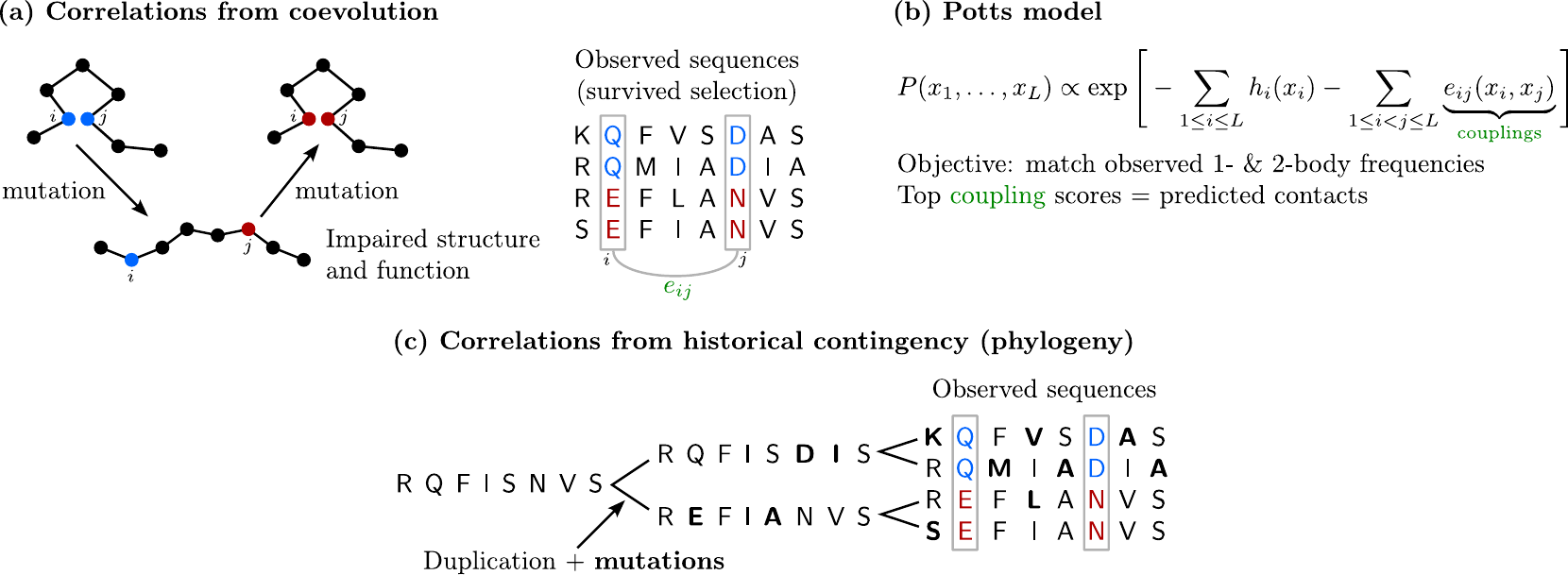}
    \caption{\textbf{Correlations from coevolution and from phylogeny in MSAs.} \textbf{(a)} Natural selection on structure and function leads to correlations between residue positions in MSAs (coevolution).
    \textbf{(b)} Potts models, also known as DCA, aim to capture these correlations in their pairwise couplings.
    \textbf{(c)} Historical contingency can lead to correlations even in the absence of structural or functional constraints. }
    \label{fig:coevolution_potts_phylogeny}
\end{figure}

MSA Transformer is known to capture three-dimensional contacts through its (tied) row attention heads~\cite{rao2021msa}, and we have shown that it also captures Hamming distances, and thus phylogeny, through its column attention heads. 
Correlations observed between the columns of an MSA can arise both from coevolution due to functional constraints and from phylogeny (see \cref{fig:coevolution_potts_phylogeny}). 
How efficiently does MSA transformer disentangle correlations from contacts and phylogeny?
We address this question in the concrete case of structure prediction.
Because correlations from contacts and phylogeny are always both present in natural data, we constructed controlled synthetic data by sampling from Potts models (\cref{fig:coevolution_potts_phylogeny}(b)), either independently at equilibrium, or along a phylogenetic tree inferred from the natural MSA using FastTree \cite{Price10}.
The Potts models we used were trained on each of 15 full natural MSAs (see \qnameref{subsec:datasets} and \cref{tab:MSA}) using the generative method bmDCA~\cite{Figliuzzi18,Russ20} -- see \qnameref{subsec:methods_phylogeny}.
This setup allows us to compare data where all correlations come from couplings (pure Potts model) to data that comprises phylogenetic correlations on top of these couplings.
For simplicity, let us call ``contacts'' the top scoring pairs of amino-acid sites according to the bmDCA models used to generate our MSAs, and refer to the task of inferring these top scoring pairs as ``contact prediction''.

Contact maps inferred by plmDCA~\cite{Ekeberg13,Ekeberg14} and by MSA Transformer for our synthetic datasets are shown in \cref{fig:contact_prediction_phylo}.
For datasets generated with phylogeny, more false positives, scattered across the whole contact maps, appear in the inference by plmDCA than in that by MSA Transformer.
This is shown quantitatively in \cref{tab:AUC}, which reports the area under the receiver operating characteristic curve (ROC-AUC) for contact prediction for two different cutoffs on the number of contacts.
We also quantify the degradation in performance caused by phylogeny by computing the relative drop $\Delta$ in ROC-AUC due to the injection of phylogeny in our generative process, for each Pfam family and for both plmDCA and MSA Transformer.
On average, $\Delta$ is twice or three times (depending on the cutoff) higher for plmDCA than for MSA Transformer.
We checked that these outcomes are robust to changes in the strategy used to compute plmDCA scores.
In particular, the average $\Delta$ for plmDCA becomes even larger when we average scores coming from independent models fitted on the $10$ subsampled MSAs used for MSA Transformer -- thus using the exact same method as for predicting contacts with MSA Transformer (see \qnameref{par:methods_generation_phylogeny}).
The conclusion is the same if $10$ (or $6$, for Pfam family PF13354) twice-deeper subsampled MSAs are employed. 

\begin{table}[htb]
    \centering
    \begin{tabular}{@{}llllclllclllclll@{}}
        \toprule
         & \multicolumn{7}{c}{ROC-AUC for $N$ contacts} &\vphantom{}& \multicolumn{7}{c}{ROC-AUC for $2L$ contacts} \\
         \cmidrule{2-8} \cmidrule{10-16}
         & \multicolumn{3}{c}{plmDCA} &\vphantom{}& \multicolumn{3}{c}{MSA Trans.} && \multicolumn{3}{c}{plmDCA} &\vphantom{}& \multicolumn{3}{c}{MSA Trans.} \\
        \cmidrule{2-4} \cmidrule{6-8} \cmidrule{10-12} \cmidrule{14-16}
        \multicolumn{1}{c}{Pfam ID} & \multicolumn{1}{c}{Eq.} & \multicolumn{1}{c}{Tree} & \multicolumn{1}{c}{$\Delta$} && \multicolumn{1}{c}{Eq.} & \multicolumn{1}{c}{Tree} & \multicolumn{1}{c}{$\Delta$} && \multicolumn{1}{c}{Eq.} & \multicolumn{1}{c}{Tree} & \multicolumn{1}{c}{$\Delta$} && \multicolumn{1}{c}{Eq.} & \multicolumn{1}{c}{Tree} & \multicolumn{1}{c}{$\Delta$} \\
        \midrule
        PF00004 & 0.87 & 0.58 & 0.33 && 0.70 & 0.67 & 0.04 && 0.93 & 0.61 & 0.34 && 0.80 & 0.71 & 0.11 \\
        PF00005 & 0.93 & 0.67 & 0.28 && 0.79 & 0.76 & 0.03 && 0.96 & 0.74 & 0.23 && 0.81 & 0.82 & $-0.01$ \\
        PF00041 & 0.86 & 0.64 & 0.25 && 0.69 & 0.62 & 0.10 && 0.94 & 0.73 & 0.22 && 0.87 & 0.79 & 0.09 \\
        PF00072 & 0.94 & 0.73 & 0.23 && 0.86 & 0.77 & 0.10 && 0.99 & 0.85 & 0.14 && 0.94 & 0.87 & 0.08 \\
        PF00076 & 0.92 & 0.69 & 0.25 && 0.81 & 0.76 & 0.05 && 0.97 & 0.72 & 0.25 && 0.88 & 0.83 & 0.05 \\
        PF00096 & 0.88 & 0.54 & 0.39 && 0.68 & 0.54 & 0.21 && 0.92 & 0.54 & 0.41 && 0.78 & 0.54 & 0.30 \\
        PF00153 & 0.95 & 0.71 & 0.26 && 0.83 & 0.63 & 0.24 && 0.98 & 0.77 & 0.21 && 0.90 & 0.65 & 0.28 \\
        PF00271 & 0.91 & 0.62 & 0.32 && 0.78 & 0.72 & 0.07 && 0.95 & 0.67 & 0.29 && 0.85 & 0.77 & 0.10 \\
        PF00397 & 0.85 & 0.58 & 0.33 && 0.69 & 0.58 & 0.15 && 0.93 & 0.61 & 0.34 && 0.76 & 0.59 & 0.22 \\
        PF00512 & 0.94 & 0.74 & 0.21 && 0.84 & 0.77 & 0.08 && 0.97 & 0.78 & 0.20 && 0.88 & 0.81 & 0.08 \\
        PF00595 & 0.91 & 0.61 & 0.33 && 0.72 & 0.62 & 0.14 && 0.96 & 0.64 & 0.33 && 0.83 & 0.68 & 0.18 \\
        PF01535 & 0.85 & 0.66 & 0.23 && 0.66 & 0.63 & 0.05 && 0.88 & 0.72 & 0.18 && 0.73 & 0.72 & 0.01 \\
        PF02518 & 0.93 & 0.69 & 0.27 && 0.82 & 0.75 & 0.09 && 0.98 & 0.78 & 0.20 && 0.90 & 0.79 & 0.12 \\
        PF07679 & 0.85 & 0.63 & 0.26 && 0.68 & 0.64 & 0.05 && 0.95 & 0.77 & 0.19 && 0.85 & 0.80 & 0.05 \\
        PF13354 & 0.68 & 0.56 & 0.18 && 0.76 & 0.65 & 0.14 && 0.82 & 0.65 & 0.21 && 0.91 & 0.74 & 0.19 \\
        \cmidrule{1-16} Average & 0.88 & 0.64 & 0.27 && 0.75 & 0.68 & 0.10 && 0.94 & 0.71 & 0.25 && 0.85 & 0.74 & 0.12 \\
        \bottomrule
    \end{tabular}
    \caption{\textbf{Impact of phylogeny on contact prediction by plmDCA and MSA Transformer.}
    We consider synthetic MSAs generated by sampling Potts models either at equilibrium (Eq.) or along inferred phylogenies (Tree). 
    We report the ROC-AUCs for contact prediction, computed by comparing couplings inferred from our synthetic MSAs using plmDCA and MSA Transformer, with ground-truth proxy contacts consisting of either the $N$ or the $2L$ pairs with top coupling scores according to the Potts models that generated the data (see \qnameref{subsec:methods_phylogeny}).
    Here, $N$ denotes the number of pairs of residues that have an all-atom minimal distance smaller than $\SI{8}{\angstrom}$ in the experimental structure in \cref{tab:MSA}, excluding pairs at positions $i, j$ with $| i - j | \leq 4$ (in all cases, $N>2L$).
    To assess the impact of phylogenetic noise, we compute $\Delta := (A_{\mathrm{eq}} - A_{\mathrm{tree}}) / A_{\mathrm{eq}}$, where $A_{\mathrm{eq}}$ is the ROC-AUC obtained from the equilibrium MSA and $A_{\mathrm{tree}}$ is the ROC-AUC obtained from the MSA with phylogeny.
    \label{tab:AUC}}
\end{table}

These results demonstrate that contact inference by MSA Transformer is less deteriorated by phylogenetic correlations than contact inference by DCA.
This resilience might explain the remarkable result that structural contacts are predicted more accurately by MSA Transformer than by Potts models even when MSA Transformer's pre-training dataset minimizes diversity \cite[Sec.\ 5.1]{rao2021msa}.

\cref{tab:AUC} also shows that plmDCA performs better than MSA Transformer on the synthetic MSAs generated without phylogeny. Because these sequences are sampled independently and at equilibrium from Potts models inferred from the natural MSAs, they are by definition well-described by Potts models. However, these sequences incorporate the imperfections of the inferred Potts models (see the inferred contact maps versus the experimental ones in \cref{fig:contact_prediction_plmDCA}), in addition to lacking the phylogenetic relationships that exist in natural MSAs. These differences with the natural MSAs that were used to train MSA Transformer might explain why it performs less well than plmDCA on these synthetic MSAs, while the opposite holds for natural MSAs (see Ref.~\cite{rao2021msa} and \cref{fig:contact_prediction_plmDCA,fig:contact_prediction_MSA_Transformer}). Note that directly comparing the performance of inference between natural and synthetic data is difficult because the ground-truth contacts are not the same and because synthetic data relies on inferred Potts models and inferred phylogenetic trees with their imperfections. However, this does not impair our comparisons of the synthetic datasets generated without and with phylogeny, or of plmDCA and MSA Transformer on the same datasets. Furthermore, an interesting feature that can be observed in \cref{fig:contact_prediction_phylo}, and is quantified in \cref{tab:ppv_and_md_experimental}, is that MSA Transformer tends to recover the experimental contact maps from our synthetic data generated by bmDCA.
Specifically, some secondary structure features that were partially lost in the bmDCA inference and generation process (see the experimental contact maps in \cref{fig:contact_prediction_plmDCA}) become better defined again upon contact inference by MSA Transformer.
This could be because MSA Transformer has learnt the structure of contact maps, including the spatial compactness and shapes of secondary structures.

\section*{Discussion}

MSA Transformer is known to capture structural contacts through its (tied) row attention heads~\cite{rao2021msa}.
Here, we showed that it also captures Hamming distances, and thus phylogenetic information, through its column attention heads.
This separation of the two signals in the representation of MSAs built by MSA Transformer comes directly from its architecture with interleaved row and column attention heads. 
It makes sense, given that some correlations between columns (i.e.\ amino-acid sites) of an MSA are associated to contacts between sites, while similarities between rows (i.e.\ sequences) arise from relatedness between sequences~\cite{Casari95}.
Specifically, we found that simple combinations of column attention heads, tuned to individual MSAs, can predict pairwise Hamming distances between held-out sequences with very high accuracy.
The larger coefficients in these combinations are found in early layers in the network.
More generally, this study demonstrated that the regressions trained on different MSAs had major similarities. 
This motivated us to train a single model across a heterogeneous collection of MSAs, and this general model was still found to accurately predict pairwise distances in test MSAs from entirely distinct Pfam families.
This result hints at a universal representation of phylogenetic relationships in MSA Transformer.
Furthermore, our results suggest that the network has learned to quantify phylogenetic relatedness by attending not only to dissimilarity \cite{rao2021msa}, but also to similarity relationships.

Next, to test the ability of MSA Transformer to disentangle phylogenetic correlations from functional and structural ones, we focused on unsupervised contact prediction tasks.
Using controlled synthetic data, we showed that unsupervised contact prediction is more robust to phylogeny when performed by MSA Transformer than by inferred Potts models. 

Language models often capture important properties of the training data in their internal representations~\cite{Rogers20}.
For instance, those trained on single protein sequences learn structure and binding sites~\cite{vig2021bertology}, and those trained on chemical reactions learn how atoms rearrange~\cite{Schwaller21}.
Our finding that detailed phylogenetic relationships between sequences are learnt by MSA Transformer, in addition to structural contacts, and in an orthogonal way, demonstrates how precisely this model represents the MSA data structure.
We note that, without language models, analyzing the correlations in MSAs can reveal evolutionary relatedness and sub-families~\cite{Casari95}, as well as collective modes of correlation, some of which are phylogenetic and some functional~\cite{Halabi09}.
Furthermore, Potts models capture the clustered organization of protein families in sequence space~\cite{Figliuzzi18}, and the latent space of variational autoencoder models trained on sequences~\cite{Riesselman18,Ding19,McGee21} qualitatively captures phylogeny~\cite{Ding19}.
Here, we demonstrated the stronger result that detailed pairwise phylogenetic relationships between sequences are quantitatively learnt by MSA Transformer. 

Separating coevolutionary signals encoding functional and structural constraints from phylogenetic correlations arising from historical contingency constitutes a key problem in analyzing the sequence-to-function mapping in proteins~\cite{Casari95,Halabi09}.
Phylogenetic correlations are known to obscure the identification of structural contacts by traditional coevolution methods, in particular by inferred Potts models~\cite{Weigt09,Marks11,Qin18,Vorberg18,RodriguezHorta19,RodriguezHorta21}, motivating various corrections~\cite{Lichtarge96,Dunn08,Ekeberg13,Marks11,Morcos11,Hockenberry19,Malinverni20,Colavin22}.
From a theoretical point of view, disentangling these two types of signals is a fundamentally hard problem~\cite{WeinsteinPreprint}.
In this context, the fact that protein language models such as MSA Transformer learn both signals in orthogonal representations, and separate them better than Potts model, is remarkable. 

Here, we have focused on Hamming distances as a simple measure of phylogenetic relatedness between sequences.
It would be very interesting to extend our study to other, more detailed, measures of phylogeny.
One may ask whether they are encoded in deeper layers in the network than those most involved in our study.
Besides, we have mainly considered attentions averaged over columns, but exploring in more detail the role of individual columns would be valuable, especially given the impact we found for column entropies.
More generally, our results suggest that the performance of protein language models trained on MSAs could be assessed by evaluating not only how well they capture structural contacts, but also how well they capture phylogenetic relationships.
In addition, the ability of protein language models to learn phylogeny could make them particularly well-suited at generating synthetic MSAs capturing the data distribution of natural ones~\cite{SgarbossaPreprint}. 
It also raises the question of their possible usefulness to infer phylogenies and evolutionary histories.

\section*{Methods}

\subsection*{Datasets}
\label{subsec:datasets}

The Pfam database \cite{mistry2021pfam} contains a large collection of related protein regions (families), typically associated to functional units called domains that can be found in multiple protein contexts.
For each of its families, Pfam provides an expert-curated seed alignment that contains a representative set of sequences.
In addition, Pfam provides deeper ``full'' alignments, that are automatically built by searching against a large sequence database using a profile hidden Markov model (HMM) built from the seed alignments.

For this work, we considered $15$ Pfam families, and for each we constructed (or retrieved, see below) one MSA from its seed alignment -- henceforth referred to as the ``seed MSA'' -- and one from its full alignment -- henceforth referred to as the full MSA.
The seed MSAs were created by first aligning Pfam seed alignments (Pfam version 35.0, Nov.\ 2021) to their HMMs using the \texttt{hmmalign} command from the HMMER suite (\url{http://hmmer.org}, version 3.3.2), and then removing columns containing only insertions or gaps.
We retained the original Pfam tree ordering, with sequences ordered according to phylogeny inferred by FastTree \cite{Price10}.
In the case of family PF02518, out of the initial 658 sequences, we kept only the first 500 in order to limit the memory requirements of our computational experiments to less than $64\,$GB.
Of the full MSAs, six (PF00153, PF00397, PF00512, PF01535, PF13354) were created from Pfam full alignments (Pfam version 34.0, Mar.\ 2021), removing columns containing only insertions or gaps, and finally removing sequences where $10 \%$ or more characters were gaps.
The remaining nine full MSAs were retrieved from the online repository \url{https://github.com/matteofigliuzzi/bmDCA} (publication date: Dec.\ 2017) and were previously considered in Ref.~\cite{Figliuzzi18}. These alignments were constructed from full Pfam alignments from an earlier release of Pfam.

An MSA is a matrix $\mathcal{M}$ with $L$ columns, representing the different amino-acid sites, and $M$ rows. Each row $i$, denoted by $\bm{x}^{(i)}$, represents one sequence of the alignment. We will refer to $L$ as the MSA length, and to $M$ as its depth.
For all but one (PF13354) of our full MSAs, $M > 36000$.
Despite their depth, however, our full MSAs include some highly similar sequences due to phylogenetic relatedness, a usual feature of large alignments of homologous proteins.
We computed the effective depth~\cite{Weigt09} of each MSA $\mathcal{M}$ as
\begin{equation}\label{eq:Meff}
    M^{(\delta)}_\mathrm{eff} := \sum_{i=1}^M w_i, \quad \text{with} \quad w_i := |\{ i' : d_\mathrm{H}(\bm{x}^{(i)}, \bm{x}^{(i')}) < \delta \}|^{-1},
\end{equation}
where $d_\mathrm{H}(\bm{x}, \bm{y})$ is the (normalized) Hamming distance between two sequences $\bm{x}$ and $\bm{y}$, i.e.\ the fraction of sites where the amino acids differ, and we set $\delta = 0.2$.
While $M^{(0.2)}_\mathrm{eff} / M$ can be as low as $0.06$ for our full MSAs, this ratio is close to 1 for all seed MSAs: it is almost $0.83$ for PF00004, and larger than $0.97$ for all other families.

Finally, for each Pfam domain considered, we retrieved  one experimental three-dimensional protein structure, corresponding to a sequence present in the full MSA, from the PDB (\url{https://www.rcsb.org}).
All these structures were obtained by X-ray crystallography and have R-free values between 0.13 and 0.29.
Information about our MSAs is summarized in \cref{tab:MSA}.

All these families have been previously considered in the literature and shown to contain coevolutionary signal detectable by DCA methods~\cite{Figliuzzi18}, making our experiments on contact prediction readily comparable with previous results.
While the precise choice of Pfam families is likely immaterial for our investigation of the column attention heads computed by MSA Transformer, our domains' short lengths are convenient in view of MSA Transformer's large memory footprint -- which is $O(L M^2) + O(L^2)$.

\subsection*{MSA Transformer and column attention}
\label{subsec:MSA-Tr_attn}

We used the pre-trained MSA Transformer model introduced in Ref.~\cite{rao2021msa}, retrieved from the Python Package Index as \texttt{fair-esm 0.4.0}.
We briefly recall that this model was trained, with a variant of the masked language modeling (MLM) objective \cite{devlin2019bert}, on 26 million MSAs constructed from UniRef50 clusters (March 2018 release), and contains 100 million trained parameters.
The input to the model is an MSA with $L$ columns and $M$ rows.
First, the model pre-pends a special beginning-of-sentence token to each row in the input MSA (this is common in language models inspired by the BERT architecture \cite{devlin2019bert}).
Then, each residue (or token) is embedded independently, via a learned mapping from the set of possible amino-acid/gap symbols into $\mathbb{R}^d$ ($d = 768$).
To these obtained embeddings, the model adds two kinds of learned \cite{Rives21} scalar positional encodings \cite{gehring2017convolutional}, designed to allow the model to distinguish between (a) different aligned positions (columns), and (b) between different sequence positions (rows). (Note that removing the latter kind was shown in Ref.~\cite{rao2021msa} to have only limited impact.)
The resulting collection of $M \times (L + 1)$ $d$-dimensional vectors, viewed as an $M \times (L + 1) \times d$ array, is then processed by a neural architecture consisting of $12$ layers.
Each layer is a variant of the axial attention \cite{ho2019axial} architecture, consisting of a multi-headed (12 heads) tied row attention block, followed by a multi-headed (12 heads) column attention block, and finally by a feed-forward network.
(Note that both attention blocks, and the feed-forward network, are in fact preceded by layer normalization \cite{ba2016layer}.)
The roles of row and column attention in the context of the MLM training objective are illustrated in \cref{fig:msa_transformer_and_hamming_regr_schematics}(a). Tied row attention incorporates the expectation that 3D structure should be conserved amongst sequences in an MSA; we refer the reader to \cite{rao2021msa} for technical details.
Column attention works as follows: let $X^{(l)}_{j}$ be the $M \times d$ matrix corresponding to column $j$ in the $M \times (L + 1) \times d$ array output by the row attention block in layer $l$ with $l = 1, \ldots, 12$.
At each layer $l$ and each head $h = 1, \ldots, 12$, the model learns three $d \times d$ matrices $W^{(l, h)}_{\mathrm{Q}}$, $W^{(l, h)}_{\mathrm{K}}$ and $W^{(l, h)}_{\mathrm{V}}$ (note that these matrices, \textit{mutatis mutandis}, could be of dimension $d \times d'$ with $d' \neq d$), used to obtain three $M \times d$ matrices
\begin{equation}\label{eq:query_key_values}
    Q^{(l, h)}_{j} = X^{(l)}_j W^{(l, h)}_{\mathrm{Q}}, \quad K^{(l, h)}_{j} = X^{(l)}_j W^{(l, h)}_{\mathrm{K}} , \quad V^{(l, h)}_{j} = X^{(l)}_{j} W^{(l, h)}_{\mathrm{V}},
\end{equation}
whose rows are referred to as ``query'', ``key'', and ``value'' vectors respectively.
The column attention from MSA column $j \in \{0, \ldots, L\}$ (where $j=0$ corresponds to the beginning-of-sentence token), at layer $l$, and from head $h$, is then the $M \times M$ matrix 
\begin{equation}\label{eq:col_attn}
    A^{(l, h)}_{j} := \mathrm{softmax}_{\mathrm{row}}\left(\frac{Q^{(l, h)}_{j} {K^{(l, h)}_{j}}^\mathrm{T}}{\sqrt{d}} \right),
\end{equation}
where we denote by $\mathrm{softmax}_{\mathrm{row}}$ the application of $\mathrm{softmax}(\xi_1, \ldots \xi_d) = (e^{\xi_{1}}, \ldots, e^{\xi_{d}}) / \sum_{k=1}^d e^{\xi_{k}}$ to each row of a matrix independently, and by $(\cdot)^\mathrm{T}$ matrix transposition. 
As in the standard Transformer architecture \cite{Vaswani17}, these attention matrices are then used to compute $M \times d$ matrices $Z^{(l, h)}_{j} = A^{(l, h)}_{j} V^{(l,h)}_{j}$, one for each MSA column $j$ and head $h$.
Projecting the concatenation $Z^{(l, 1)}_{j} \mid \cdots \mid Z^{(l, 12)}_{j}$, a single $M \times d$ matrix $Z^{(l)}_{j}$ is finally obtained at layer $l$.
The collection $(Z^{(l)}_j)_{j=1, \ldots, L}$, thought of as an $M \times (L + 1) \times d$ array, is then passed along to the feed-forward layer.

\subsection*{Supervised prediction of Hamming distances}
\label{subsec:methods_supervised_hamming}

Row $i$ of the column attention matrices $A^{(l, h)}_{j}$ in \cref{eq:col_attn} consists of $M$ positive weights summing to one -- one weight per row index $i'$ in the original MSA.
According to the usual interpretation of the attention mechanism \cite{Bahdanau14, Vaswani17}, the role of these weights may be described as follows: When constructing a new internal representation (at layer $l$) for the row-$i$, column-$j$ residue position, the network distributes its focus, according to these weights, among the $M$ available representation vectors associated with each MSA row-$i'$, column-$j$ residue position (including $i' = i$).
Since row attention precedes column attention in the MSA Transformer architecture, we remark that, even at the first layer, the row-$i'$, column-$j$ representation vectors that are processed by that layer's column attention block can encode information about the entire row $i'$ in the MSA.

In~\cite[Sec.\ 5.1]{rao2021msa}, it was shown that, for some layers $l$ and heads $h$, averaging the $M \times M$ column attention matrices $A^{(l, h)}_{j}$ in \cref{eq:col_attn} from all MSA columns $j$, and then averaging the result along the first dimension, yields $M$-dimensional vectors whose entries correlate reasonably well with the phylogenetic sequence weights $w_i$ defined in \cref{eq:Meff}.
Larger weights are, by definition, associated with less redundant sequences, and MSA diversity is known to be important for coevolution-based methods -- particularly in structure prediction tasks.
Thus, these correlations can be interpreted as suggesting that the model is, in part, explicitly attending to a maximally diverse set of sequences.

Beyond this, we hypothesize that MSA Transformer may have learned to quantify and exploit phylogenetic correlations in order to optimize its performance in the MLM training objective of filling in randomly masked residue positions.
To investigate this, we set up regression tasks in which, to predict the Hamming distance $y$ between the $i$-th and the $i'$-th sequence in an MSA $\mathcal{M}$ of length $L$, we used the entries $a^{(l, h)}_{i, i'}$ at position $(i, i')$ (henceforth $a^{(l, h)}$ for brevity) from the $144$ matrices
\begin{equation}
    \label{eq:A^lh}
    \bm{A}^{(l, h)} := \frac{1}{2(L + 1)}\sum_{j = 0}^L \left(A^{(l, h)}_{j} + {A^{(l, h)}_{j}}^{\mathrm{T}} \right), \quad \text{with} \quad 1 \leq l \leq 12 \text{ and } 1 \leq h \leq 12.
\end{equation}
These matrices are obtained by averaging, across all columns $j = 0, \ldots, L$, the symmetrised column attention maps $A^{(l, h)}_{j}$ computed by MSA Transformer, when taking $\mathcal{M}$ as input.
We highlight that column $j = 0$, corresponding to the beginning-of-sentence token, is included in the average defining $\bm{A}^{(l, h)}$.

We fit fractional logit models via quasi-maximum likelihood estimation \cite{papke1996econometric} using the \texttt{statsmodels} package  (version 0.13.2)  \cite{seabold2010statsmodels}.
Namely, we model the relationship between the Hamming distance $y$ and the aforementioned symmetrised, and averaged, attention values $\bm{a} = (a^{(1,1)}, \ldots, a^{(12,12)})$, as
\begin{equation}
    \label{eq:logistic_model_defn}
    \mathbb{E}[y \, | \, \bm{a}] = G_{\beta_0, \, \bm{\beta}}(\bm{a}), \quad \text{with} \quad G_{\beta_0, \, \bm{\beta}}(\bm{a}) := \sigma\left( \beta_0 + \bm{a} \bm{\beta}^{\mathrm{T}} \right),
\end{equation}
where $\mathbb{E}[ \, \cdot \, | \, \cdot \, ]$ denotes conditional expectation, $\sigma(x) = (1 + e^{-x})^{-1}$ is the standard logistic function, and the coefficients $\beta_0$ and $\bm{\beta} = (\beta_1, \ldots, \beta_{144})$ are determined by maximising the sum of Bernoulli log-likelihoods
\begin{equation}
    \label{eq:logistic_model_loss}
    \ell(\beta_0, \bm{\beta} \, | \, \bm{a}, y) = y \log[G_{\beta_0, \,  \bm{\beta}}(\bm{a})] + (1 - y) \log[1 - G_{\beta_0, \, \bm{\beta}}(\bm{a})],
\end{equation}
evaluated over a training set of observations of $y$ and $\bm{a}$.
Note that this setup is similar to logistic regression, but allows for the dependent variable to take real values between $0$ and $1$ (it can be equivalently described as a generalized linear model with binomial family and logit link).
For simplicity, we refer to these fractional logit models simply as ``logistic models''.
Our general approach to predict Hamming distances is illustrated in \cref{fig:msa_transformer_and_hamming_regr_schematics}(b).

Using data from our seed MSAs (cf.\ \cref{tab:MSA}), we performed two types of regression tasks.
In the first one, we randomly partitioned the set of row indices in each separate MSA $\mathcal{M}$ into two subsets $I_{\mathcal{M}, \, \mathrm{train}}$ and $I_{\mathcal{M}, \, \mathrm{test}}$, with $I_{\mathcal{M}, \, \mathrm{train}}$ containing $70\%$ of the indices.
We then trained and evaluated one model for each $\mathcal{M}$, using as training data the Hamming distances, and column attentions, coming from (unordered) pairs of indices in $I_{\mathcal{M}, \, \mathrm{train}}$, and as test data the Hamming distances, and column attentions, coming from pairs of indices in $I_{\mathcal{M}, \, \mathrm{test}}$.
The second type of regression task was a single model fit over a training dataset consisting of all pairwise Hamming distances, and column attentions, from the first 12 of our 15 MSAs.
We then evaluated this second model over a test set constructed in an analogous way from the remaining 3 MSAs.

\subsection*{Synthetic MSA generation via Potts model sampling along inferred phylogenies} \label{subsec:methods_phylogeny}

To assess the performance of MSA Transformer at disentangling signals encoding functional and structural (i.e.\ fitness) constraints from phylogenetic correlations arising from historical contingency, we generated and studied controlled synthetic data.
Indeed, disentangling fitness landscapes from phylogenetic history in natural data poses a fundamental challenge \cite{WeinsteinPreprint} -- see \cref{fig:coevolution_potts_phylogeny} for a schematic illustration.
This makes it very difficult to assess the performance of a method at this task directly on natural data, because gold standards where the two signals are well-separated are lacking. 
We resolved this conundrum by generating synthetic MSAs according to well-defined dynamics such that the presence of phylogeny can be controlled.

First, we inferred unrooted phylogenetic trees from our full MSAs (see \qnameref{subsec:datasets}), using FastTree version 2.1 \cite{Price10} with its default settings.
Our use of FastTree is motivated by the depth of the full MSAs, which makes it computationally prohibitive to employ more precise inference methods.
Deep MSAs are needed for the analysis described below, since it relies on accurately fitting Potts models.

Then, we fitted Potts models on each of these MSAs using bmDCA \cite{Figliuzzi18} (\url{https://github.com/ranganathanlab/bmDCA}, version 0.8.12) with its default hyperparameters.
These include, in particular, regularization strengths for the Potts model fields and couplings, both set at $\lambda = 10^{-2}$.
With the exception of family PF13354, we trained all models for $2000$ iterations and stored the fields and couplings at the last iteration; in the case of PF13354, we terminated training after $1480$ iterations.
In all cases, we verified that, during training, the model's loss had converged.
The choice of bmDCA is motivated by the fact that, as has been shown in Refs.~\cite{Figliuzzi18, Russ20}, model fitting on natural MSAs using Boltzmann machine learning yields Potts models with good generative power.
This sets it apart from other DCA inference methods, especially pseudo-likelihood DCA (plmDCA)~\cite{Ekeberg13, Ekeberg14}, which is the DCA standard for contact prediction, but cannot faithfully reproduce empirical one- and two-body marginals, making it a poor choice of a generative model~\cite{Figliuzzi18}.

Using the phylogenetic trees and Potts models inferred from each full MSA, we generated synthetic MSAs without or with phylogeny, as we now explain.
In the remainder of this subsection, let $\mathcal{M}$ denote an arbitrary MSA from our set of full MSAs, $L$ its length, and $M$ its depth.

Consider a sequence of $L$ amino-acid sites.
We denote by $x_i \in \{1, \dots, q\}$ the state of site $i \in \{1, \dots, L\}$, where $q=21$ is the number of possible states, namely the 20 natural amino acids and the alignment gap.
A general Potts model Hamiltonian applied to a sequence $\bm{x} = (x_1, \ldots, x_{L})$ reads
\begin{equation}
    H(\bm{x}) = -\sum_{i=1}^{L}h_{i}(x_i) - \sum_{j=1}^{L}\sum_{i=1}^{j-1}e_{ij}(x_i, x_j)\,,
\label{eq:maxent}
\end{equation}
where the fields $h_i(x_i)$ and couplings $e_{ij}(x_i,x_j)$ are parameters that can be inferred from data by DCA methods~\cite{Weigt09,Cocco18}.
In our case, they are inferred from $\mathcal{M}$ by bmDCA~\cite{Figliuzzi18, Russ20}.
The Potts model probability distribution is then given by the Boltzmann distribution associated to the Hamiltonian $H$ in \cref{eq:maxent}:
\begin{equation}
    P(\bm{x})=\frac{e^{-H(\bm{x})}}{Z}\,,
\end{equation}
where $Z$ is a constant ensuring normalization.
In this context, we implement a Metropolis--Hastings algorithm for Markov Chain Monte Carlo (MCMC) sampling from $P$, where an iteration step consists of a proposed move (mutation) in which a site $i$ is chosen uniformly at random, and its state $x_i$ may be changed into another state chosen uniformly at random.
Each of these attempted mutations is accepted or rejected according to the Metropolis criterion, i.e.\ with probability
\begin{equation}
    p = \min \left[ 1, \exp \left(-\Delta H \right)\right],
    \label{eq:MetroCrit}
\end{equation}
where $\Delta H$ is the difference in the value of $H$ after and before the mutation.

\paragraph{Generating independent equilibrium sequences under a Potts model.}
To generate a synthetic MSAs without phylogeny from each $\mathcal{M}$, we performed equilibrium MCMC sampling from the Potts model with Hamiltonian $H$ in \cref{eq:maxent}, using the Metropolis--Hastings algorithm.
Namely, we started from a set of $M$ randomly and independently initialized sequences, and proposed a total number $N$ of mutations on each sequence.
Suitable values for $N$ are estimated by bmDCA during its training, to ensure that Metropolis--Hastings sampling reaches thermal equilibrium after $N$ steps when starting from a randomly initialized sequence~\cite{Figliuzzi18}.
We thus used the value of $N$ estimated by bmDCA at the end of training.
This yielded a synthetic MSA of the same depth $M$ as the original full MSA $\mathcal{M}$, composed of independent equilibrium sequences.

\paragraph{Generating sequences along an inferred phylogeny under a Potts model.}\label{par:methods_generation_phylogeny}
We also generated synthetic data using MCMC sampling along our inferred phylogenetic trees~\cite{Vorberg18}, using an open-source implementation available at \url{https://github.com/Bitbol-Lab/Phylogeny-Partners} (version 2.0).
We started from an equilibrium ancestor sequence sampled as explained above, and placed it at the root (note that, while FastTree roots its trees arbitrarily, root placement does not matter; see below).
Then, this sequence was evolved by successive duplication (at each branching of the tree) and mutation events (along each branch).
Mutations were again modeled using for acceptance the Metropolis criterion in \cref{eq:MetroCrit} with the Hamiltonian in \cref{eq:maxent}.
As the length $b$ of a branch gives the estimated number of substitutions that occurred per site along it~\cite{Price10}, we generate data by making a number of accepted mutations on this branch equal to the integer closest to $b L$.
Since we traversed the entire inferred tree in this manner, the resulting sequences at the leaves of the tree yield a synthetic MSA of the same depth as the original full MSA $\mathcal{M}$.
Finally, we verified that the Hamming distances between sequences in these synthetic MSAs were reasonably correlated with those between corresponding sequences in the natural MSAs -- see \cref{fig:Hamming_synthetic_MSAs}.

Because we start from an ancestral equilibrium sequence, and then employ the Metropolis criterion, all sequences in the phylogeny are equilibrium sequences.
Thus, some of the correlations between the sequences at the leaves of the tree can be ascribed to the couplings in the Potts model, as in the case of independent equilibrium sequences described above.
However, their relatedness adds extra correlations, arising from the historical contingency in their phylogeny.
Note that separating these ingredients is extremely tricky in natural data~\cite{WeinsteinPreprint}, which motivates our study of synthetic data.

Our procedure for generating MSAs along a phylogeny is independent of the placement of the tree's root.
Indeed, informally, a tree's root placement determines the direction of evolution; hence, root placement should not matter when evolution is a time-reversible process.
That evolution via our mutations and duplications is a time-reversible process is a consequence of the fact that we begin with equilibrium sequences at the (arbitrarily chosen) root.
More formally, for an irreducible Markov chain with transition matrix $\mathcal{P}$ and state space $\Omega$, and for any $n \geq 1$, let $\mathrm{Markov}_n(\pi, \mathcal{P})$ denote the probability space of chains $(X_k)_{0 \leq k \leq n}$ with initial distribution $\pi$ on $\Omega$.
If $\pi$ is the chain's stationary distribution and $\pi$ satisfies detailed balance, then, for any number of steps $n \geq 1$, any chain $(X_k)_{0 \leq k \leq n} \in \mathrm{Markov}_n(\pi, \mathcal{P})$ is reversible in the sense that $(X_{n - k})_{0 \leq k \leq n} \in \mathrm{Markov}_n(\pi, \mathcal{P})$.
In our case, since the Metropolis--Hastings algorithm constructs an irreducible Markov chain whose stationary distribution satisfies detailed balance, and since duplication events are also time-reversible constraints imposed at each branching node, all ensemble observables are independent of root placement as long as the root sequences are sampled from the stationary distribution.

\paragraph{Assessing performance degradation due to phylogeny in coupling inference.}\label{par:performance_degradation}
DCA methods and MSA Transformer both offer ways to perform unsupervised inference of structural contacts from MSAs of natural proteins.
In the case of DCA, the established methodology~\cite{Ekeberg13,Ekeberg14,Figliuzzi18} is to (1) learn fields and couplings [see \cref{eq:maxent}] by fitting the Potts model, (2) change the gauge to the zero-sum gauge, (3) compute the Frobenius norms, for all pairs of sites $(i,j)$, of the coupling matrices $\left( e_{ij}(x, y) \right)_{x, y}$, and finally (4) apply the average product correction (APC)~\cite{Dunn08}, yielding a coupling score $E_{ij}$.
Top scoring pairs of sites are then predicted as being contacts.
In the case of MSA Transformer~\cite{rao2021msa}, a single logistic regression (shared across all possible input MSAs) was trained to regress contact maps from a sparse linear combination of the symmetrized and APC-corrected row attention heads (see \qnameref{subsec:MSA-Tr_attn}).

We applied these inference techniques, normally used to predict structural contacts, on our synthetic MSAs generated without and with phylogeny (see above).
As proxies for structural contacts, we used the pairs of sites with top coupling scores in the Potts models used to generate the MSAs.
Indeed, when presented with our synthetic MSAs generated at equilibrium, DCA methods for fitting Potts models should recover the ranks of these coupling scores well.
Hence, their performance in this task provide a meaningful baseline against which performance when a phylogeny was used to generate the data, as well as MSA Transformer's performance, can be measured.

As a DCA method to infer these coupling scores, we used plmDCA~\cite{Ekeberg13,Ekeberg14} as implemented in the \texttt{PlmDCA} Julia package (\url{https://github.com/pagnani/PlmDCA}, version 0.4.1), which is the state-of-the-art DCA method for contact inference.
We fitted one plmDCA model per synthetic MSA, using default hyperparameters throughout; these include, in particular, regularization strengths set at $\lambda = 10^{-2}$ for both fields and couplings, and automatic estimation of the phylogenetic cutoff $\delta$ in \cref{eq:Meff}.
We verified that these settings led to good inference of structural contacts on the original full MSAs by comparing them to the PDB structures in \cref{tab:MSA} -- see \cref{fig:contact_prediction_plmDCA}.
For each synthetic MSA, we computed coupling scores $E_{ij}$ for all pairs of sites.

While Potts models need to be fitted on deep MSAs to achieve good contact prediction, MSA Transformer's memory requirements are considerable even at inference time, and the average depth of the MSAs used to train MSA Transformer was 1192~\cite{rao2021msa}.
Concordantly, we could not run MSA Transformer on any of the synthetic MSAs in their entirety.
Instead, we subsampled each synthetic MSA 10 times, by selecting each time a number $M_\mathrm{sub}$ of row indices uniformly at random, without replacement.
We used $M_\mathrm{sub} \approx 380$ for family PF13354 due to its greater length, and $M_\mathrm{sub} \approx 500$ for all other families.
Then, we computed for each subsample a matrix of coupling scores using MSA Transformer's row attention heads and the estimated contact probabilities from the aforementioned logistic regression.
Finally, we averaged the resulting 10 matrices to obtain a single matrix of coupling scores.
We used a similar strategy (and the same randomly sampled row indices) to infer structural contact scores from the natural MSAs -- see \cref{fig:contact_prediction_MSA_Transformer}.
Consistently with findings in Ref.~\cite{rao2021msa}, MSA Transformer generally performs better than plmDCA (\cref{fig:contact_prediction_plmDCA}) at contact inference.

\section*{Data availability}
All datasets used in our work (namely, the natural MSAs and PDB structures described in \qnameref{subsec:datasets} and listed in \cref{tab:MSA}, as well as the corresponding synthetic MSAs generated for the analysis in \qnameref{subsec:methods_phylogeny}) are available at \url{https://zenodo.org/record/7096792}.

\section*{Code availability}
Our code is available at \url{https://zenodo.org/record/7096792}.

\bibliographystyle{naturemag} 
\bibliography{bibliography} 

\section*{Acknowledgments}
The authors thank Mohammed AlQuraishi for inspiring discussions, Sergey Ovchinnikov for valuable feedback on the first version of this manuscript, Tom Sercu for clarifying some aspects of MSA Transformer's pre-training, and Alexander Rives for interesting conversations.
Code by Andonis Gerardos was used to generate synthetic MSAs along prescribed phylogenies.
This project has received funding from the European Research Council (ERC) under the European Union’s Horizon 2020 research and innovation programme (grant agreement No.~851173, to A.-F.~B.).

\section*{Author contributions statement}
All authors participated in the design of the project. U.\ L.\ and D.\ S.\ wrote the software. U.\ L.\ built the datasets and performed the bmDCA analysis. U.\ L.\ and D.\ S.\ performed the analysis of MSA Transformer column attention. All authors interpreted the results. A.-F.\ B.\ supervised the project. U.\ L.\ and A.-F.\ B.\ wrote the manuscript. All authors reviewed and edited the manuscript.

\section*{Competing interests statement}
The authors declare no competing interests.

\newpage

\begin{center}
 {\LARGE \bf Supplementary material}   
\end{center}

\renewcommand{\thesection}{S\arabic{section}}
\renewcommand{\figurename}{Supplementary Figure}
\setcounter{figure}{0}
\renewcommand{\tablename}{Supplementary Table}
\setcounter{table}{0} 

\section*{Supplementary tables}

\begin{table}[h!]
    \centering
    \begin{tabular}{@{}llllclllcll@{}}
        \toprule
        & & \multicolumn{2}{c}{Seed MSA} & & \multicolumn{3}{c}{Full MSA} & &  \multicolumn{2}{c}{PDB structure} \\
        \cmidrule{3-4} \cmidrule{6-8} \cmidrule{10-11}
        \multicolumn{1}{c}{Pfam ID} & \multicolumn{1}{c}{Family name} & \multicolumn{1}{c}{$L$} & \multicolumn{1}{c}{$M$} && \multicolumn{1}{c}{$L$} & \multicolumn{1}{c}{$M$} & \multicolumn{1}{c}{$M_\mathrm{eff}^{(0.2)}$} && \multicolumn{1}{c}{ID} & \multicolumn{1}{c}{Resol.} \\
        \midrule
        PF00004 & AAA & 132 & 207 && 132 & 39277 & 9050 && 4D81 & $\SI{2.40}{\angstrom}$ \\
        PF00005 & ABC\_tran & 137 & 55 && 137 & 68891 & 43882 && 1L7V & $\SI{3.20}{\angstrom}$ \\
        PF00041 & fn3 & 85 & 98 && 85 & 42721 & 17783 && 3UP1 & $\SI{2.15}{\angstrom}$  \\
        PF00072 & Response\_reg & 112 & 52 && 112 & 73063 & 40180 && 3ILH & $\SI{2.59}{\angstrom}$ \\
        PF00076 & RRM\_1 & 68 & 70 && 69 & 51964 & 20276 && 3NNH & $\SI{2.75}{\angstrom}$  \\
        PF00096 & zf-C2H2 & 23 & 159 && 23 & 38996 & 12581 && 4R2A & $\SI{1.59}{\angstrom}$ \\
        PF00153 & Mito\_carr & 97 & 160 && 94 & 93776 & 17860 && 1OCK & $\SI{2.20}{\angstrom}$ \\
        PF00271 & Helicase\_C & 111 & 421 && 111 & 66809 & 25018 && 3EX7 & $\SI{2.30}{\angstrom}$ \\
        PF00397 & WW & 31 & 448 && 31 & 39045 & 3361 && 4REX & $\SI{1.60}{\angstrom}$ \\
        PF00512 & HisKA & 67 & 265 && 66 & 154998 & 67303 && 3DGE & $\SI{2.80}{\angstrom}$ \\
        PF00595 & PDZ & 82 & 44 && 82 & 71303 & 4053 && 1BE9 & $\SI{1.82}{\angstrom}$ \\
        PF01535 & PPR & 31 & 458 && 31 & 109064 & 37514 && 4M57 & $\SI{2.86}{\angstrom}$ \\
        PF02518 & HATPase\_c & 112 & 500 && 111 & 80714 & 59190 && 3G7E & $\SI{2.20}{\angstrom}$ \\
        PF07679 & I-set & 90 & 48 && 90 & 36141 & 14611 && 1FHG & $\SI{2.00}{\angstrom}$ \\
        PF13354 & Beta-lactamase2 & 215 & 76 && 198 & 4642 & 3535 && 6QW8 & $\SI{1.10}{\angstrom}$ \\
        \bottomrule
    \end{tabular}
    \caption{\textbf{Pfam families and MSAs considered in this work.} For each family, we considered a shallow MSA constructed from the corresponding Pfam seed alignment (Seed MSA) and a deep MSA constructed from the Pfam full alignment (Full MSA) -- see \qnameref{subsec:datasets}. For both kinds of MSA, we report the length $L$ and depth $M$. For the full MSAs, we report the effective depth $M^{(0.2)}_\mathrm{eff}$ as defined in \cref{eq:Meff}. The PDB structures used, and their resolutions, are also reported. Note that occasional length mismatches between seed and full MSAs reflect our use of data from a more recent Pfam release in the case of the seed MSAs. \label[stab]{tab:MSA}}
\end{table}

\newpage

\begin{table}[h!]
    \centering
    \begin{tabular}{llll}
        \toprule
        \multicolumn{1}{c}{Family} & \multicolumn{1}{c}{$R^2$} & \multicolumn{1}{c}{Pearson} & \multicolumn{1}{c}{Slope} \\
        \midrule
        PF00004 & 0.84 & 0.95 & 0.95 \\
        PF00005 & 0.72 & 0.92 & 0.75 \\
        PF00041 & 0.56 & 0.90 & 0.75 \\
        PF00072 & 0.66 & 0.90 & 0.71 \\
        PF00076 & 0.59 & 0.88 & 0.68 \\
        PF00096 & 0.57 & 0.88 & 0.73 \\
        PF00153 & 0.81 & 0.93 & 0.80 \\
        PF00271 & 0.77 & 0.93 & 1.11 \\
        PF00397 & 0.23 & 0.84 & 1.13 \\
        PF00512 & 0.77 & 0.93 & 0.94 \\
        PF00595 & 0.50 & 0.89 & 0.63 \\
        PF01535 & 0.54 & 0.86 & 1.18 \\
        \textbf{PF02518} & \textbf{0.60} & \textbf{0.90} & \textbf{1.20} \\
        \textbf{PF07679} & \textbf{0.28} & \textbf{0.85} & \textbf{0.57} \\
        \textbf{PF13354} & \textbf{0.67} & \textbf{0.92} & \textbf{0.70} \\
        \bottomrule
    \end{tabular}
    \caption{\textbf{Quality of fit for our logistic model trained on Hamming distances and column attentions from several MSAs.}
    For the logistic model described in \qnameref{subsec:results_common_hamming}, and for the MSAs in the training set (plain font) and test set (boldface font), we report (1) the $R^2$ coefficient of determination for the model's predictions, (2) the Pearson correlation coefficient between predictions and ground truth Hamming distances, and (3) the slope of the line of best fit when regressing the ground truth Hamming distances on the model's predictions.\label[stab]{tab:mean_cols_regr_results_many_msas_fit_quality}}
\end{table}

\begin{table}[h!]
    \centering
    \begin{tabular}{llll}
        \toprule
        \multicolumn{1}{c}{Family} & \multicolumn{1}{c}{$R^2$} & \multicolumn{1}{c}{Pearson} & \multicolumn{1}{c}{Slope} \\
        \midrule
        PF00004 & 0.31 & 0.67 & 1.60 \\
        PF00005 & 0.33 & 0.59 & 0.96 \\
        PF00041 & 0.02 & 0.43 & 1.02 \\
        PF00072 & 0.45 & 0.67 & 1.07 \\
        PF00076 & 0.30 & 0.56 & 1.10 \\
        PF00096 & -0.13 & 0.24 & 0.52 \\
        PF00153 & 0.07 & 0.32 & 0.66 \\
        PF00271 & 0.13 & 0.46 & 1.25 \\
        PF00397 & -0.31 & 0.39 & 1.19 \\
        PF00512 & 0.02 & 0.35 & 0.73 \\
        PF00595 & 0.49 & 0.70 & 1.09 \\
        PF01535 & -0.17 & 0.20 & 0.63 \\
        \textbf{PF02518} & \textbf{-0.09} & \textbf{0.32} & \textbf{1.07} \\
        \textbf{PF07679} & \textbf{0.29} & \textbf{0.57} & \textbf{0.91} \\
        \textbf{PF13354} & \textbf{0.35} & \textbf{0.62} & \textbf{1.15} \\
        \bottomrule
    \end{tabular}
    \caption{\textbf{Quality of fit for a logistic model trained on Hamming distances and column attentions from several MSAs, using MSA Transformer with random weights.}
    We reinitialized MSA Transformer's parameters to random values, using the same protocols originally used in pre-training (see \qnameref{subsec:results_common_hamming}).
    Results for the same task as in \cref{tab:mean_cols_regr_results_many_msas_fit_quality} are shown.\label[stab]{tab:mean_cols_regr_results_many_msas_random_init_fit_quality}}
\end{table}

\newpage

\begin{table}[h!]
    \centering
    \begin{tabular}{lccc}
        \toprule
        Family  &  Pearson  &  Z-score \\
        \midrule
        PF00004 & 0.36 & -2.29 \\
        PF00005 & 0.46 & -2.78 \\
        PF00041 & 0.38 & -2.21 \\
        PF00072 & 0.41 & -1.60 \\
        PF00076 & 0.53 & -1.64 \\
        PF00096 & 0.88 & -1.71 \\
        PF00153 & 0.66 & -3.06 \\
        PF00271 & 0.83 & -2.24 \\
        PF00397 & 0.79 & -1.21 \\
        PF00512 & 0.66 & -1.73 \\
        PF00595 & 0.13 & 0.28 \\
        PF01535 & 0.89 & -1.68 \\
        PF02518 & 0.81 & -2.48 \\
        PF07679 & 0.66 & -2.20 \\
        PF13354 & 0.39 & -2.96 \\
        \bottomrule
    \end{tabular}
    \caption{\textbf{Relation between entropy and test error when using parameters from our common logistic model on individual column attention matrices.}
    We applied the common logistic model described in \qnameref{subsec:results_common_hamming} -- which was trained on the column-wise means of MSA Transformer's column attention heads [see \cref{eq:A^lh}] -- to individual column attention heads.
    We then computed the resulting errors in the prediction of Hamming distances. 
    For each of our seed MSAs, we report the Pearson correlation between the entropy of each column and the standard deviation of the distribution of these errors.
    We also computed the mean of this standard deviation when restricting to the 5 columns with lowest entropy.
    Z-scores for this mean are reported, showing that these 5 columns have significantly lower standard deviations than the rest (except for PF00595).
    \label[stab]{tab:low_entropy_table}}
\end{table}

\newpage

\begin{table}[h!]
    \centering
    \begin{tabular}{@{}lllcllcllcll@{}}
        \toprule
         & \multicolumn{5}{c}{PPV} &\vphantom{}& \multicolumn{5}{c}{Median distance (\AA)} \\
         \cmidrule{2-6} \cmidrule{8-12}
         & \multicolumn{2}{c}{plmDCA} &\vphantom{}& \multicolumn{2}{c}{MSA Trans.} && \multicolumn{2}{c}{plmDCA} &\vphantom{}& \multicolumn{2}{c}{MSA Trans.} \\
        \cmidrule{2-3} \cmidrule{5-6} \cmidrule{8-9} \cmidrule{11-12}
        \multicolumn{1}{c}{Pfam ID} & \multicolumn{1}{c}{Eq.} & \multicolumn{1}{c}{Tree} && \multicolumn{1}{c}{Eq.} & \multicolumn{1}{c}{Tree} && \multicolumn{1}{c}{Eq.} & \multicolumn{1}{c}{Tree} && \multicolumn{1}{c}{Eq.} & \multicolumn{1}{c}{Tree} \\
        \midrule
        PF00004 & 0.27 & 0.06 && 0.50 & 0.14 && 8.5 & 13.8 && 5.2 & 9.1 \\
        PF00005 & 0.27 & 0.11 && 0.44 & 0.18 && 8.4 & 15.3 && 6.5 & 8.8 \\
        PF00041 & 0.33 & 0.16 && 0.45 & 0.35 && 7.3 & 12.1 && 5.2 & 7.0 \\
        PF00072 & 0.43 & 0.21 && 0.56 & 0.32 && 6.1 & 8.6 && 4.7 & 6.8 \\
        PF00076 & 0.45 & 0.25 && 0.61 & 0.42 && 6.4 & 9.4 && 4.4 & 6.7 \\
        PF00153 & 0.22 & 0.14 && 0.30 & 0.23 && 10.1 & 12.5 && 8.6 & 10.2 \\
        PF00271 & 0.30 & 0.14 && 0.50 & 0.16 && 7.7 & 10.5 && 5.1 & 7.8 \\
        PF00397 & 0.40 & 0.21 && 0.61 & 0.27 && 6.0 & 10.6 && 3.7 & 8.5 \\
        PF00512 & 0.24 & 0.18 && 0.34 & 0.32 && 9.5 & 11.1 && 7.9 & 7.8 \\
        PF00595 & 0.35 & 0.16 && 0.60 & 0.29 && 6.8 & 10.0 && 4.2 & 8.3 \\
        PF01535 & 0.37 & 0.23 && 0.45 & 0.34 && 7.5 & 9.2 && 7.1 & 7.6 \\
        PF02518 & 0.25 & 0.15 && 0.33 & 0.30 && 9.2 & 11.2 && 7.7 & 8.9 \\
        PF07679 & 0.42 & 0.23 && 0.54 & 0.33 && 4.3 & 8.5 && 3.9 & 6.2 \\
        PF13354 & 0.17 & 0.07 && 0.52 & 0.17 && 11.0 & 15.2 && 3.7 & 7.6 \\
        \cmidrule{1-12} Average & 0.32 & 0.17 && 0.48 & 0.27 && 7.8 & 11.3 && 5.6 & 7.9 \\
        \bottomrule
    \end{tabular}
    \caption{\textbf{Comparing predicted contact maps using plmDCA and MSA Transformer on synthetic MSAs with experimental structures, for $\bm{14}$ of our full MSAs.}
    For each of the Pfam families in \cref{fig:contact_prediction_phylo} we obtained experimental contact maps, using the PDB structures in \cref{tab:MSA}, by selecting the $2L$ closest residue pairs at positions $i, j$ with $| i - j | \geq 5$.
    Using these structural contacts as ground truths, we computed the positive predictive values (PPVs) of the $2L$ top-scoring pairs, according to plmDCA or MSA Transformer, when performing contact inference on the synthetic MSAs generated by our Potts models either without phylogeny (Equilibrium) or with phylogeny (Tree) -- see \qnameref{subsec:methods_phylogeny}.
    We also report the median across predicted pairs of the inter-residue distance (all-atom minimal distance) in the reference experimental 3D structure (Median distance).
    \label[stab]{tab:ppv_and_md_experimental}}
\end{table}

\newpage
\section*{Supplementary figures}

\begin{figure}[h!]
    \centering
    \begin{subfigure}[h]{0.5\textwidth}
        \centering
        \includegraphics[width=\textwidth]{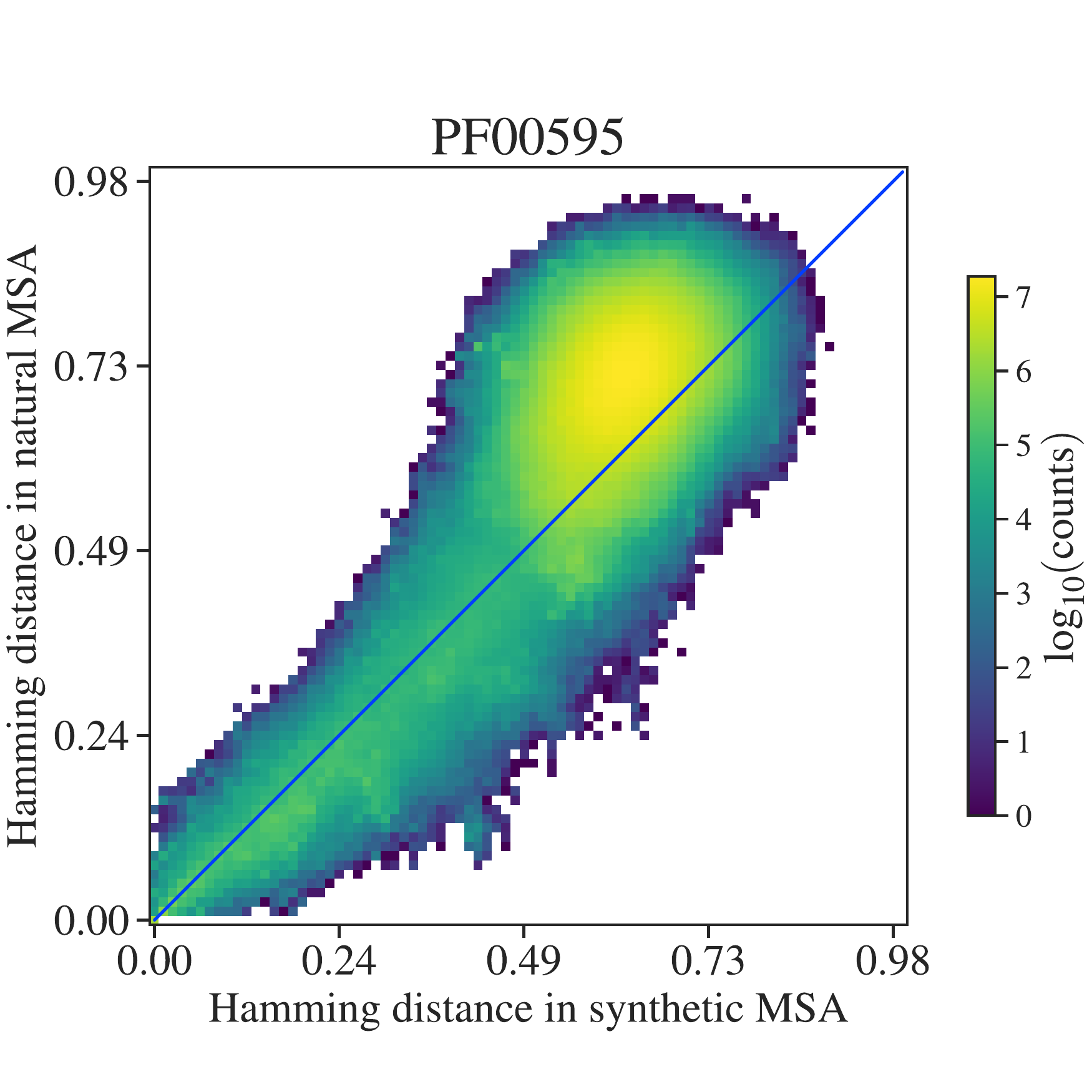}
        \caption{\label{subfig:Hamming_hist2D_PF00595_synthetic_vs_natural}}
    \end{subfigure}
    \hfill
    \begin{subfigure}[h]{0.35\textwidth}
        \centering
        \begin{tabular}{ccc}
            \toprule
             & \multicolumn{2}{c}{Pearson correlation} \\
            \cmidrule{2-3} Family & All & $d_{\mathrm{nat}} \leq 0.5$ \\
            \midrule
            PF00004 & 0.72 & 0.75 \\
            PF00005 & 0.31 & 0.62 \\
            PF00041 & 0.31 & 0.90 \\
            PF00072 & 0.33 & 0.85 \\
            PF00076 & 0.25 & 0.84 \\
            PF00096 & 0.32 & 0.23 \\
            PF00153 & 0.39 & 0.91 \\
            PF00271 & 0.37 & 0.70 \\
            PF00397 & 0.60 & 0.73 \\
            PF00512 & 0.30 & 0.65 \\
            PF00595 & 0.67 & 0.96 \\
            PF01535 & 0.15 & 0.70 \\
            PF02518 & 0.39 & 0.67 \\
            PF07679 & 0.19 & 0.90 \\
            PF13354 & 0.74 & 0.79 \\
            \bottomrule
        \end{tabular}
        \caption{\label{subfig:Pearson_Hamming_synthetic_MSAs}}
    \end{subfigure}
    \caption{\textbf{Comparing Hamming distances from synthetic MSAs generated along inferred phylogenies with Hamming distances from natural MSAs.}
    The Hamming distances between sequences in our synthetic MSAs generated along phylogenies (see \qnameref{par:methods_generation_phylogeny}) are reasonably correlated with those between corresponding sequences in the natural MSAs used to infer the phylogenies.
    \subref{subfig:Hamming_hist2D_PF00595_synthetic_vs_natural} Density plot comparing the Hamming distances between sequences in the synthetic MSA corresponding to family PF00595, with those between corresponding sequences in the natural MSA for this family.
    The pixel with coordinates $(d_\mathrm{synth}, d_\mathrm{nat})$ is colored according to the base-$10$ logarithm of the number of pairs of indices $(i, j)$ such that the Hamming distance between row $i$ and row $j$ in the synthetic MSA is $d_\mathrm{synth}$, and the distance between the same rows in the natural MSA is $d_\mathrm{nat}$.
    Recall that the sequence at row $i$ of the synthetic MSA has been generated on the leaf of the inferred phylogenetic tree that corresponds to the natural sequence at row $i$ of the natural MSA.
    As is visible here, our synthetic MSAs tend to have slightly smaller Hamming distances, on average, than their natural counterparts.
    \subref{subfig:Pearson_Hamming_synthetic_MSAs} For each Pfam family, we computed two Pearson correlation coefficients: first, the correlation between all Hamming distances among synthetic sequences and all Hamming distances among natural sequences; second, the same correlation but restricting to pairs of indices $(i, j)$ such that the distance between sequence $i$ and sequence $j$ in the natural MSA is no larger than $0.5$.}
    \label[sfig]{fig:Hamming_synthetic_MSAs}
\end{figure}

\newpage

\begin{figure}[h!]
    \centering
    \includegraphics[width=0.75\textwidth]{contact_prediction_plmDCA}
    \caption{\textbf{Predicted contact maps using plmDCA versus experimental contact maps for $\bm{14}$ of our full natural MSAs.}
    Experimental contact maps are displayed in the upper-triangular portions of each panel, and are obtained from the PDB structures in \cref{tab:MSA} by using an all-atom minimal Euclidean distance cutoff of $\SI{8}{\angstrom}$, excluding residue pairs at positions $i, j$ with $| i - j | \leq 4$.
    Predictions are displayed in the lower-triangular portions, and are obtained considering the top $2 L$ scores, where $L$ is the length of the MSA.
    Light blue squares represent true positive predictions, dark blue squares false negative predictions, and red squares false positive predictions.
    For each predicted contact map, we report the positive predictive value (PPV) given these choices.
    Results for PF00096 are not displayed due to its very short length.
    \label[sfig]{fig:contact_prediction_plmDCA}}
\end{figure}

\newpage

\begin{figure}[h!]
    \centering
    \includegraphics[width=0.75\textwidth]{contact_prediction_MSA_Transformer}
    \caption{\textbf{Predicted contact maps using MSA Transformer versus experimental contact maps for $\bm{14}$ of our full natural MSAs.}
    Same as in \cref{fig:contact_prediction_plmDCA}, but using MSA Transformer instead of plmDCA to infer structural contacts.
    Contact scores were computed for each natural MSA as follows: (1) the MSA was subsampled 10 times using the same randomly sampled row indices used for the corresponding synthetic MSAs (see \qnameref{par:performance_degradation}); (2) for each subsample, a matrix of contact scores was computed using MSA Transformer's row attention heads and the estimated contact probabilities from the logistic regression trained in \cite{rao2021msa}; (3) the resulting 10 matrices were averaged to obtain a single matrix of contact scores.
    \label[sfig]{fig:contact_prediction_MSA_Transformer}}
\end{figure}

\newpage

\begin{figure}[h!]
    \centering
    \includegraphics[width=0.87\textwidth]{contact_prediction_phylo_1_nearestint_v2}
    \vspace{-1em}
    \caption{(Continued on the next page.) \label[sfig]{fig:contact_prediction_phylo}}
\end{figure}

\newpage

\begin{figure}[h!]\ContinuedFloat
    \centering
    \includegraphics[width=0.87\textwidth]{contact_prediction_phylo_2_nearestint_v2}
    \vspace{-1em}
    \caption{(Continued on the next page.)}
\end{figure}

\newpage

\begin{figure}[h!]\ContinuedFloat
    \caption{\textbf{Predicted contact maps using plmDCA and MSA Transformer on synthetic MSAs, versus ground truth proxy contact maps defined by top Potts model couplings, for $\bm{14}$ of our full MSAs.}
    Contact maps containing $2L$ contacts, and obtained from the ground-truth couplings in the Potts models used to generate our synthetic MSAs (see \qnameref{subsec:methods_phylogeny}), are displayed in the upper-triangular portions of each panel.
    In the lower-triangular portions, we display the $2L$ top-scoring pairs according to plmDCA or MSA Transformer, when performing contact inference on synthetic MSAs generated from those Potts models either without phylogeny (Equilibrium) or with phylogeny (Tree).
    Light blue squares represent true positive predictions, dark blue squares false negative predictions, and red squares false positive predictions.
    For each predicted contact map, we report the positive predictive value (PPV) given these choices.
    ROC-AUC values corresponding to this analysis are reported in the columns titled ``ROC-AUC for $2L$ contacts'' of \cref{tab:AUC}.
    \vspace{19cm}}
\end{figure}

\end{document}